\documentclass[a4paper,11pt]{article}
\usepackage{pos}
\usepackage{graphicx}
\usepackage{caption}
\usepackage{amsmath}
\usepackage{subcaption}
\usepackage{hyperref}
\usepackage{mathtools}
\usepackage{float}
\usepackage{rotating}
\usepackage{url}
\usepackage[font=scriptsize,labelfont=bf,skip=6pt]{caption}
\usepackage[T1]{fontenc}

\title{Installation of proANUBIS -- a proof-of-concept demonstrator for the ANUBIS experiment}

\author*[a]{Aashaq Shah}

\author[]{\textnormal{for the ANUBIS collaboration}}

\affiliation[a]{Department of Physics, Cavendish Laboratory, University of Cambridge\\
  J J Thomson Avenue, Cambridge, CB3 0HE, United Kingdom}

\emailAdd{aashaq.shah@cern.ch}

\abstract{AN Underground Belayed In-Shaft search experiment (ANUBIS) was proposed to search for neutral long-lived particles (LLP's) at CERN's ATLAS underground cavern. A prototype or a proof-of-concept demonstrator detector — proANUBIS was recently installed to prove the feasibility of such an experiment. The prototype demonstrator is expected to play a role in validating simulation studies and providing insights into the anticipated backgrounds for the ANUBIS experiment. The current report provides an overview of the experimental setup for this prototype detector, and its  commissioning and installation details.}

\FullConference{The Eleventh Annual Conference on Large Hadron Collider Physics (LHCP2023)\\
 22-26 May 2023\\
 Belgrade, Serbia\\}

\begin{document}
\maketitle

\section{Introduction}
\begin{figure}[H]
    \centering
    \begin{subfigure}[b]{0.85\textwidth}
        \includegraphics[width=4.8cm, height=6.3cm]{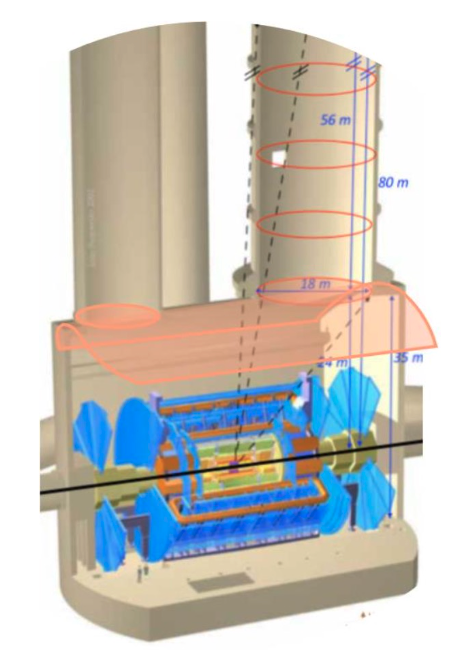} 
        \includegraphics[width=7.5cm, height=5.4cm]{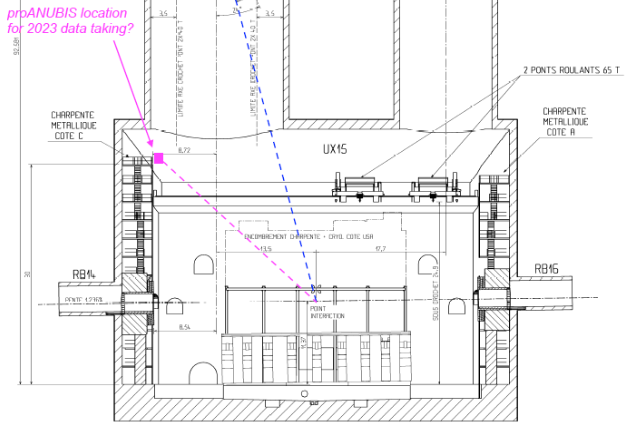} 
    \end{subfigure}
    \caption{(left) The sketch depicts the layout of the underground cavern at LHC Point 1, featuring the ATLAS experiment. Additionally, PX14 access shafts with the four ANUBIS stations as proposed originally are shown alongside the PX16 shaft. The ATLAS underground cavern ceiling highlighted with orange colour illustrates the proposed configuration of the ANUBIS experiment. Inset within the sketch are two disks covering the shafts are going to be an integral part of the ANUBIS experiment. (right) The sketch depicting the location of the proANUBIS setup, the location is on the level 12 floor side-A of the ATLAS underground cavern and the detector is facing the collision interaction point.} \label{fig:UndergroundCavern}
\end{figure}

\vspace{-10pt}
The ANUBIS experiment~\cite{ANUBIS_proposal} was proposed to complement and extend the exciting off-axis LLP search programme and its central idea is to reduce the civil engineering costs by repurposing the existing infrastructure of the ATLAS experiment at the LHC~\cite{ATLAS_experiment}. The combination of ANUBIS's large detector volume (Figure~\ref{fig:UndergroundCavern}) and the adjacency to the ATLAS experiment gives a larger acceptance compared to other proposals for a significant range of lifetimes, and the possibility to synchronize and fully integrate ANUBIS with ATLAS. The original proposal also proposed installing a prototype detector — proANUBIS at CERN LHC to demonstrate the feasibility of the proposal which is discussed in Section~\ref{Sec:proANUBIS_setup}. 
\vspace{-5pt}
\vspace{-10pt}
The aim of installing proANUBIS serves as a step towards validating background simulations. It also provides an opportunity to assess the detector's capabilities and functionalities in the challenging environment of the ATLAS underground cavern. By testing the detector's response to background signals and potential sources of interference, we hope to get first-hand experience in realising the  ANUBIS project. Also, by examining the performance of the demonstrator, we are expecting to gain insights into the mechanical engineering requirements, detector design, readout implementations and further developments of the ANUBIS detector. 
\vspace{-10pt}
\section{The proANUBIS Setup} \label{Sec:proANUBIS_setup}
\vspace{-10pt}
\subsection{Detector Design and Geometry}
\begin{figure}[H]
    \centering
    \begin{subfigure}[b]{0.85\textwidth}
        \includegraphics[width=5.7cm, height=5.6cm]{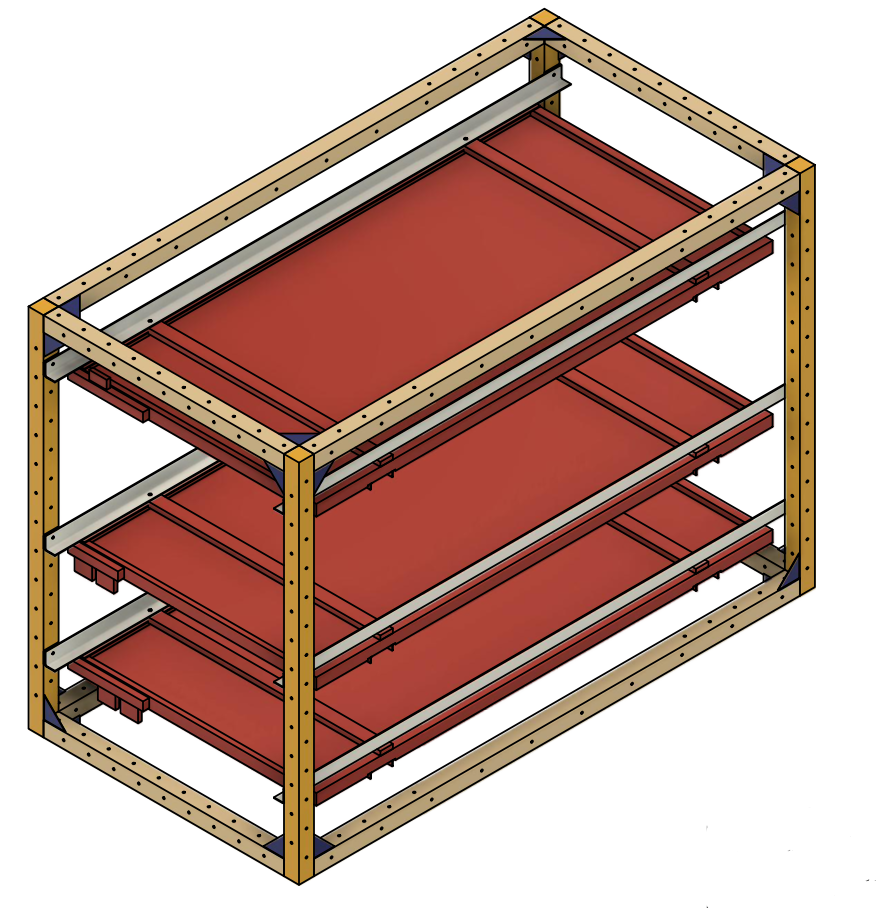} 
        \includegraphics[width=4cm, height=5.3cm]{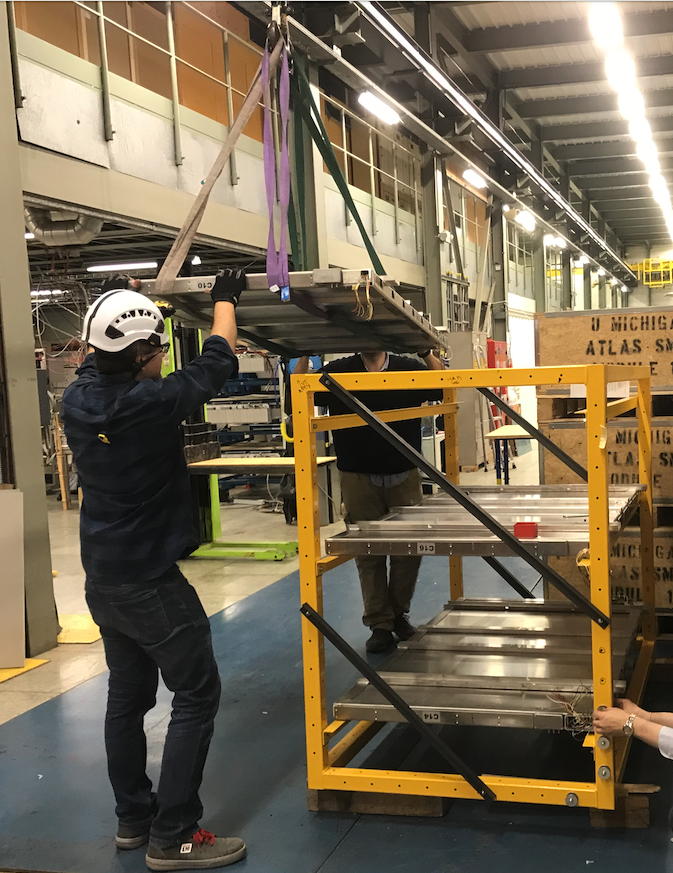} 
    \end{subfigure}
    \caption{(left) The design of the proANUBIS demonstrator is depicted, showcasing the positioning of the three integrated tracking layers in a metallic frame, (right) A picture showing proANUBIS metallic frame (in yellow colour) being populated with RPC integrated chambers with a crane at CERN BB5.} \label{fig:proANUBIS_setup}
\end{figure}

\vspace{-5pt}
The proANUBIS detector design and geometry are illustrated in Figure~\ref{fig:proANUBIS_setup}. The detector comprises three layers of tracking detectors, each measuring an area of 1 m $\times$ 2 m and the top and bottom layers being separated by approximmately half a meter with respect to the middle layer. The bottom layer consists of three individual tracking layers arranged as a triplet, while the top layer consists of two layers referred to as a doublet, and the middle layer features a single layer, aptly named the singlet, as depicted in the same Figure. The choice of having a triplet and doublet configuration is driven by the need to effectively form muon hits resulting from the LHC collisions. While these layers alone would suffice for muon tracking, the inclusion of the singlet layer proves instrumental in detecting and tracking these particles as it will provide an additional hit and will help in vertexing. This arrangement enhances the detector's capabilities, enabling a more comprehensive study of potential background signals. The spacing between individual layers is optimized to ensure efficient particle tracking within the detector volume. The overall geometry of proANUBIS is engineered to strike a balance between performance and practicality, taking into account the constraints imposed by the ATLAS underground cavern geometry/environment.
\vspace{-5pt}
\vspace{-10pt}
\subsection{Detector Technology}
\vspace{-5pt}
The proANUBIS employs detector technology based on the next generation of Resistive Plate Chambers (RPCs), known as BIS78 (Barrel Inner Short 7 and 8)~\cite{RPCUpgrade_BIS78, ATLAS_TDAQ_phase2, ATLAS_Muon_TDR_phase2}. These detectors have been recently deployed for the ATLAS BIS78 project, serving as a pilot for the Phase-II Barrel Inner (BI) RPC upgrade. They feature bakelite electrodes with a reduced gas gap of 1 mm, a departure from the standard 2 mm gap used in traditional RPC technology. These detectors are equipped with copper strip readout and feature on-detector Front-End (FE) electronics~\cite{BIS78_electronics}. One of the primary reasons for considering the BIS78 technology is its cost-effectiveness compared to other detector technologies. As ANUBIS requires a larger detector area/volume, the BIS78 technology aligns perfectly with the project's aspirations. Moreover, this technology provides good timing and spatial resolutions, approximately 250 ps and 1 mm, respectively, making it a perfect choice for studying the lifetime of LLPs.

The integration of these detectors within aluminium structures follows three combinations; a triplet of RPC chambers, doublet, and a singlet. The aluminium structures not only provide mechanical strength to the detectors but also act as ``Faraday cages'', shielding sensitive Front-End electronics against external electromagnetic interference. The integration process replaced empty spaces in the singlet and doublet configurations with aluminium honeycomb structures of 12 mm to prevent any detector movement. The connections to gas, high and low voltages were facilitated through rigid aluminium boxes. Post full integration and assembly, the leakage current and efficiency measurements were re-evaluated to ensure the integrity of the detector’s performance.
\vspace{-10pt}
\subsection{Installation at CERN ATLAS underground cavern}
\begin{figure}[H]
    \centering
    \begin{subfigure}[b]{0.85\textwidth}
        \includegraphics[width=5.4cm, height=4.2cm]{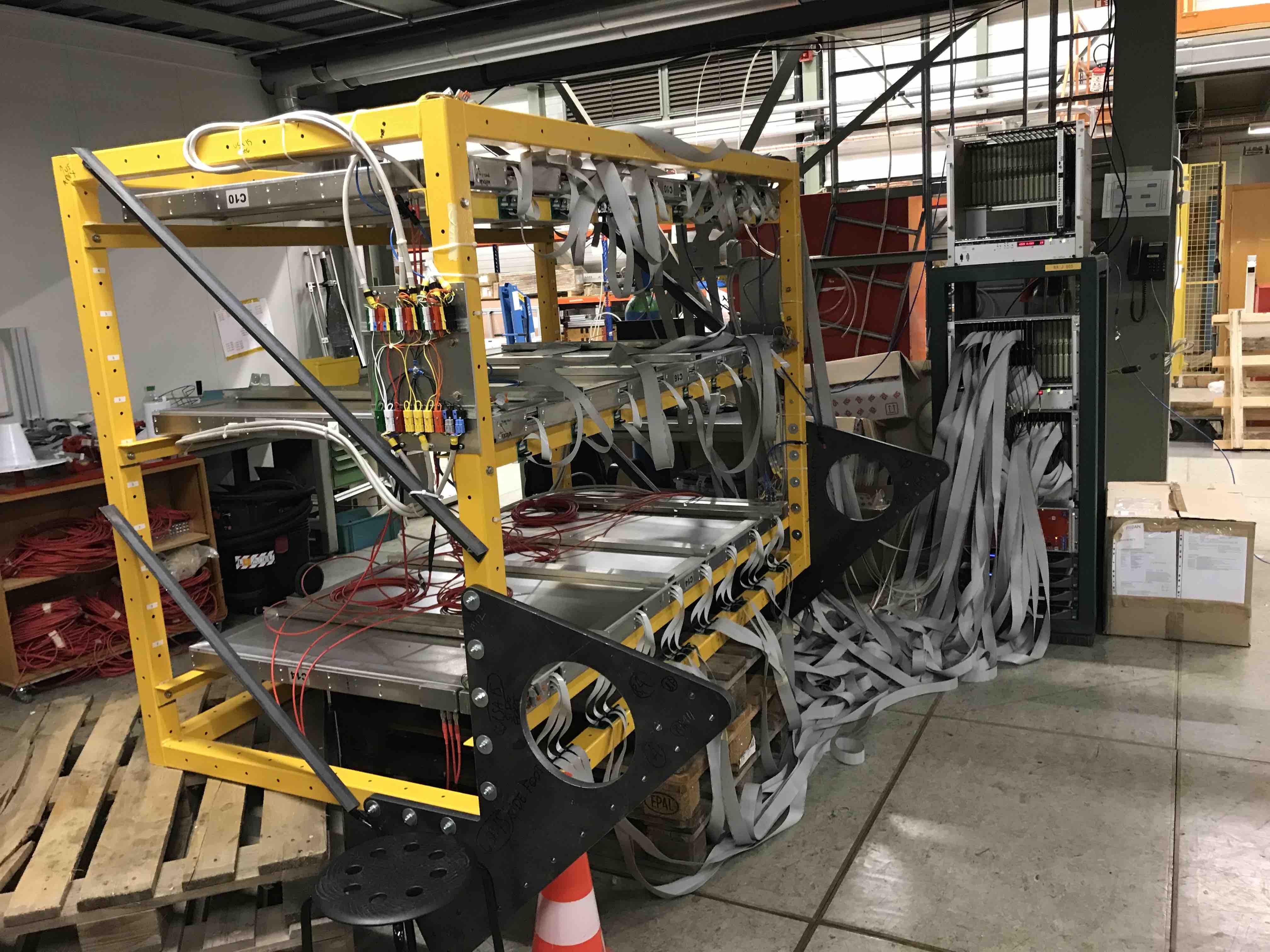} 
        \includegraphics[width=3.4cm, height=4.2cm]{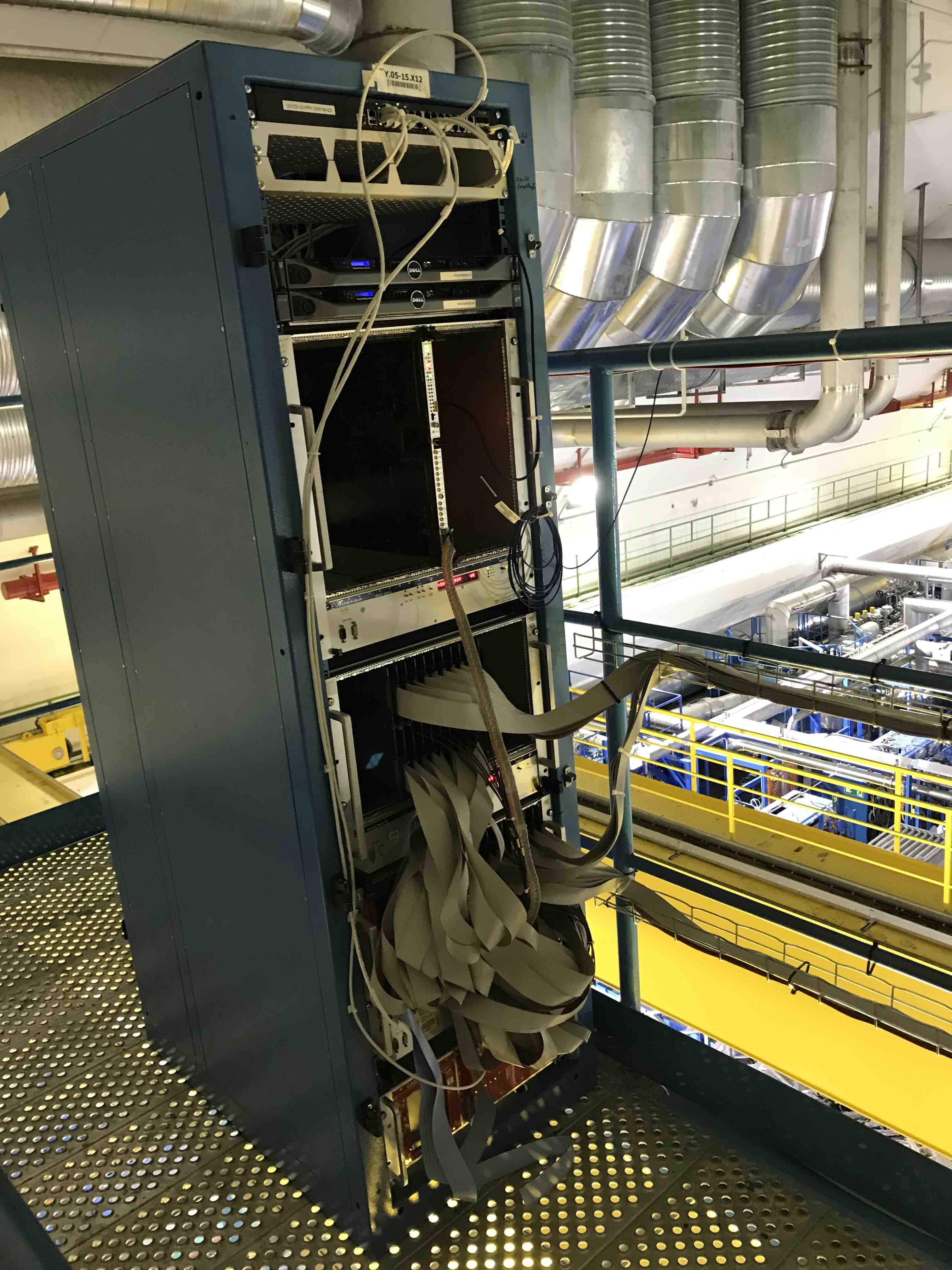} 
        \includegraphics[width=3.6cm, height=4.2cm]{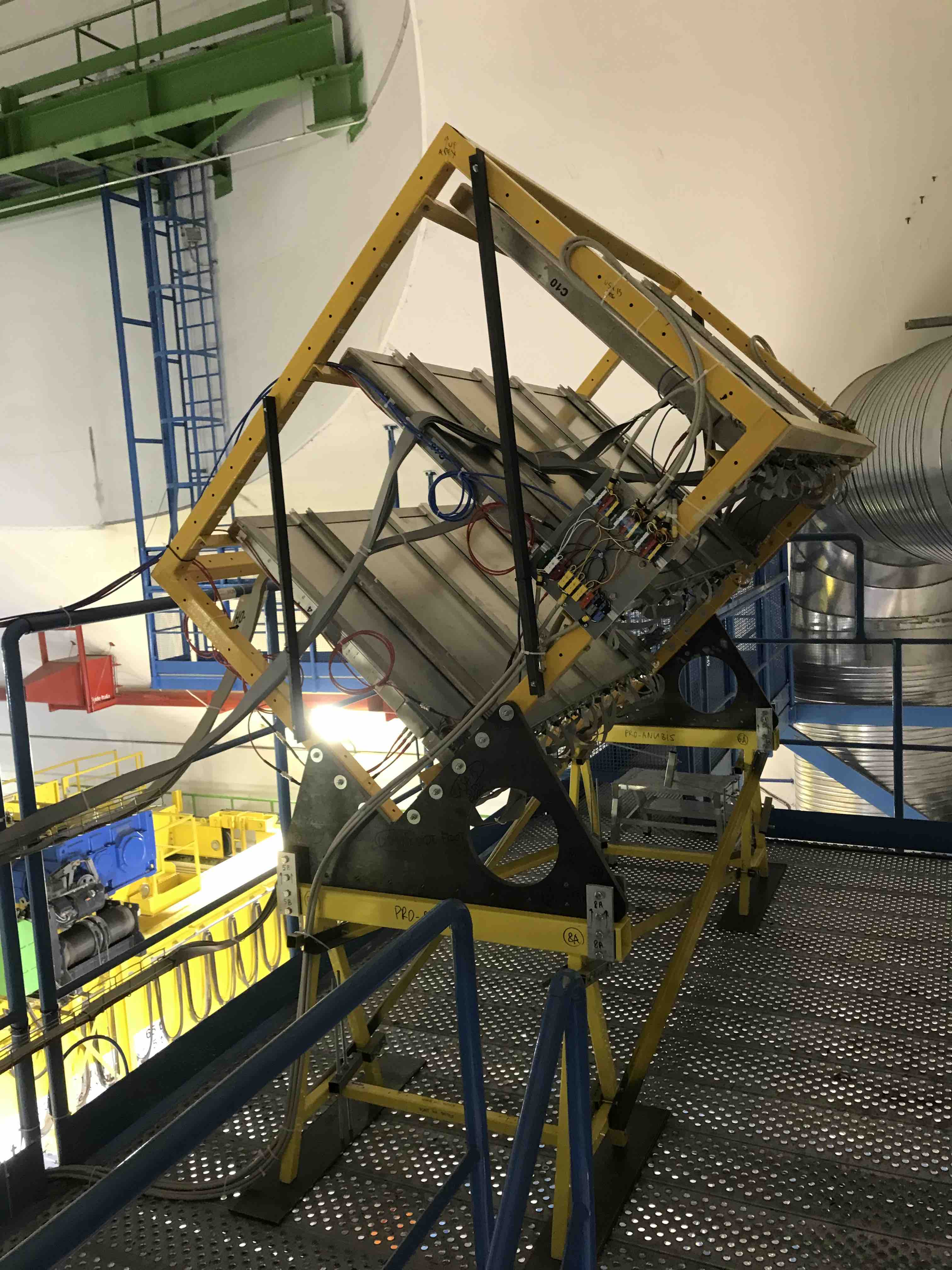} 
    \end{subfigure}
    \caption{(left) A picture showing the proANUBIS detectors and DAQ chain being tested in CERN BB5 laboratory. (middle) Picture showcasing Data Acquisition (DAQ) rack along with the proANUBIS setup (right) installed on level-12 side-A of the ATLAS underground cavern after lowering through the ATLAS PX14 access shaft.} \label{fig:Installation}
\end{figure}

\vspace{-5pt}
The commissioning of proANUBIS commenced at CERN BB5 (Figure~\ref{fig:Installation} (left)) during the summer of 2022, where integration and testing were conducted. The triplet (from bottom), singlet, and doublet RPC chambers were inserted into the metallic frame (yellow colour as can be seen in Figure~\ref{fig:proANUBIS_setup}), following the design layout depicted in Figure~\ref{fig:proANUBIS_setup}. The Data Acquisition System (DAQ) which also includes the high and low voltage systems, underwent testing at BB5 to ensure the functionalities. Once each component was examined and deemed ready for deployment, the entire proANUBIS setup was transported to its designated location at LHC Point 1 within the ATLAS experiment at CERN. To safeguard the detector's integrity during transport, diligent care was taken in handling and securing the proANUBIS unit.  
\vspace{-3pt}

\vspace{-5pt}
The installation site for proANUBIS is situated on level 12 side-A of the ATLAS cavern (Figure~\ref{fig:Installation} (right)), approximately 80 m meters below the Earth's surface. With the support of ATLAS technical coordination, the proANUBIS detector, along with the DAQ rack, was lowered into position through the PX14 access shafts, employing a crane for precise manoeuvring. The coordinated efforts helped in successfully installing the proANUBIS system, side by side with the DAQ rack, within the cavern, as showcased in Figure~\ref{fig:Installation}. The setup was connected to a low and high-voltage system, the standard RPC gas mixture was supplied to chambers from the ATLAS muon system, and the system was also connected to the LHC clock to receive the bunch crossing information. These tasks were executed during the Year End Technical Stop (YETS) of 2022, ensuring minimal disruption to the operation of the current LHC experiments.
\vspace{-10pt}
\section{Summary and outlook}
\vspace{-5pt}
The successful commissioning and installation of the proANUBIS demonstrator at the ATLAS experiment's underground cavern represents a very important step towards realising the ANUBIS experiment. The initial testing of the RPC tracking chambers and Data Acquisition (DAQ) system with cosmic rays started at the CERN BB5 laboratory. Though testing was limited due to the LHC deadlines, it still provided insights into the functionality of the detector and DAQ chain. With the prototype now installed, the experiment is well-positioned to conduct background and other studies including optimizing detector performance, validating the readout system, and fostering fruitful collaboration among various institutions.  


\end{document}